\documentclass[a4paper,10pt,twoside]{cpc-hepnp}

\usepackage{multicol}
\usepackage{graphicx}
\usepackage{booktabs}
\usepackage{amssymb,bm,mathrsfs,bbm,amscd}
\usepackage[tbtags]{amsmath}
\usepackage{lastpage}
\def\slashchar#1{\setbox0=\hbox{$#1$}
   \dimen0=\wd0 \setbox1=\hbox{/} \dimen1=\wd1
   \ifdim\dimen0>\dimen1 \rlap{\hbox to \dimen0{\hfil/\hfil}} #1
   \else  \rlap{\hbox to \dimen1{\hfil$#1$\hfil}} / \fi}
\def\vsl{\slashchar{v}}

\begin{document}

\fancyhead[co]{\footnotesize }

\footnotetext[0]{Received 14 December 2009}

\title{ Semileptonic $bc$ to $cc$  and $bb$ to $bc$ Baryon Decays and Heavy Quark Spin
 Symmetry\thanks{This work is partly supported by DGI and FEDER funds,
 under contracts FIS2006-03438, FIS2008-01143/FIS and PIE-CSIC
 200850I238. }}

\author{%
      J. Nieves$^{1)}$\email{jmnieves@ific.uv.es}%
\quad J.M. Flynn$^{2)}$\email{j.m.flynn@soton.ac.uk}%
\quad E. Hern\'andez$^{3)}$\email{gajatee@usal.es}  %
}
\maketitle

\address{%
1~Instituto de F\'isica corpuscular (IFIC), Centro Mixto
Universidad de Valencia-CSIC, Institutos de Investigaci\'on de
Paterna, Aptdo. 22085, 46071, Valencia, Spain\\
2~School of Physics and Astronomy, University of
  Southampton, Highfield, Southampton SO17~1BJ, UK\\
3~Departamento de F\'\i sica Fundamental e IUFFyM, Universidad de Salamanca, 
E-37008 Salamanca, Spain\\
}

\begin{abstract}
We study the semileptonic decays of the lowest-lying $bc$ baryons to
the lowest-lying $cc$ baryons ($\Xi_{bc}^{(\prime*)}\to
\Xi_{cc}^{(*)}$ and $\Omega_{bc}^{(\prime*)}\to \Omega_{cc}^{(*)}$) ,
in the limit $m_b, m_c \gg \Lambda_\mathrm{QCD}$ and close to the zero
recoil point. The separate heavy quark spin symmetries make it
possible to describe all these decays using a single form factor.  We
also show how these constraints can be used to test the validity of
different quark model calculations. $bb$ to $bc$ baryon decays are also
discussed.
\end{abstract}



\begin{multicols}{2}

\section{Introduction}

The static theory for a system with two heavy quarks has infra-red
divergences which can be regulated by the kinetic energy term $\bar
h_Q (D^2/2 m_Q) h_Q$. This term breaks the heavy quark flavour
symmetry, but not the spin symmetry for each heavy quark flavour. The
spin symmetry is sufficient to derive relations between form factors
for decays of hadrons containing two heavy quarks in the heavy quark
limit, as was first shown in~\citep{White:1991hz}.  The consequences
of the separate spin symmetries of each of the heavy quarks for
semileptonic decays of $B_c$ mesons were worked out
in~\citep{Jenkins:1992nb}. The formalism was extended in
Ref.~\citep{Flynn:2007qt} to describe semileptonic decays of $bc$
($bb$) baryons to $cc$ ($bc$) baryons, and its predictions were
confronted with different constituent quark model calculations in
\citep{Hernandez:2007qv}.  Here, we will  review the main
findings of Refs.~\citep{Flynn:2007qt} and~\citep{Hernandez:2007qv} on
 the semileptonic decays of baryons containing two heavy quarks and
a light quark.  

According to heavy quark spin symmetry (HQSS)~\cite{Jenkins:1992nb},
in the infinite heavy quark mass limit, one can select the heavy quark
subsystem of a doubly heavy baryon to have a well defined total spin
$S_h=0,1$. In Table~\ref{tab:hh-baryons} we show the ground state
$J^\pi=\frac12^+,\frac32^+$ doubly heavy baryons classified so that
$S_h$ is well defined. Being ground states for the given quantum
numbers, a total orbital angular momentum $L=0$ is naturally
assumed. HQSS guarantees that, in the infinite heavy quark mass limit,
all baryons with the same flavour content listed in
Table~\ref{tab:hh-baryons} are degenerate, and that a unique function
describes the entire family of decays of cascade $bc$ baryons
$\Xi_{bc}$, $\Xi'_{bc}$ and $\Xi^*_{bc}$ to cascade $cc$ baryons
$\Xi_{cc}$ and $\Xi^*_{cc}$ near the zero recoil point. In this latter
kinematical region, the velocities of the initial and final baryons
are approximately the same. If the momenta of the initial $bc$ and
final $cc$ baryons are $p_\mu = m_{bc} v_\mu$ and $p'_\mu =
m_{cc}v'_\mu= m_{cc}v_\mu + k_\mu$ respectively, then $k$ will be a
small residual momentum near the zero-recoil point, and since the
final baryon is on-shell, $k\cdot v = \mathcal{O}(1/m_{cc})$ will be
suppressed. Moreover, this unique function, which describes all the
decays, satisfies a normalization condition (a consequence of vector
current conservation) at zero-recoil if the heavy quarks are
degenerate. These results can straightforwardly be applied to the
corresponding decays involving $\Omega$ baryons and also to the decays
of $bb$ baryons to $bc$ baryons. Some of these decays have also been
studied in various quark model
approaches~\citep{Sanchis-Lozano:1994vh, Guo:1998yj,
  Ebert:2004ck,Albertus:2006ya, Faessler:2009xn,Roberts:2009}, and we
will critically review  to what extent these calculations are
consistent with HQSS.

\end{multicols}

\begin{center}
\tabcaption{Quantum numbers of the baryons analyzed in this
study. $J^\pi$ is the baryon spin parity, and $S_{h}$ is the spin
of the heavy degrees of freedom, well-defined in the infinite heavy
mass limit. $l$ denotes a  $u$ or $d$ quark.}
\begin{tabular*}{170mm}{@{\extracolsep{\fill}}cccc||cccc}
\toprule
Baryon\hspace{.3cm} & Quark content & $S_h$ & $J^\pi$ &
 Baryon\hspace{.3cm} & Quark content & $S_h$ &
$J^\pi$\\ \hline
$\Xi_{cc}$ & c~c~l & 1 & 1/2$^+$ & $\Omega_{cc}$ & c~c~s & 1 &
1/2$^+$  \\ 
$\Xi_{cc}^*$ & c~c~l & 1 & 3/2$^+$ & $\Omega_{cc}^*$ & c~c~s & 1
& 3/2$^+$  \\ 
$\Xi_{bb}$ & b~b~l & 1 & 1/2$^+$ & $\Omega_{bb}$ & b~b~s & 1 &
1/2$^+$  \\ 
$\Xi_{bb}^*$ & b~b~l & 1 & 3/2$^+$ & $\Omega_{bb}^*$ & b~b~s & 1
& 3/2$^+$ \\
$\Xi_{bc}$ & b~c~l & 1 & 1/2$^+$ & $\Omega_{bc}$ & b~c~s & 1 &
1/2$^+$  \\ 
$\Xi_{bc}'$ & b~c~l & 0 & 1/2$^+$ &$\Omega_{bc}'$ & b~c~s & 0 &
1/2$^+$  \\ 
$\Xi_{bc}^*$ & b~c~l & 1 & 3/2$^+$ & $\Omega_{bc}^*$ & b~c~s & 1
& 3/2$^+$ \\ 
\bottomrule
\label{tab:hh-baryons}
\end{tabular*}
\end{center}

\begin{multicols}{2}

To end this introduction, we devote a few words to the
effects arising from the mixing of the $\Xi$ and $\Xi^\prime$
$bc-$states\cite{Roberts:2007ni, Albertus:2009ww} (see also the talk
by E. Hen\'nandez\cite{Albertus:2009dw}) . Owing to the finite value of
the heavy quark masses, the hyperfine interaction between the light
quark and any of the heavy quarks can admix both $S_h=0$ and $S_h=1$
spin components into the wave function. This mixing should be
negligible for $bb$ and $cc$ doubly heavy baryons as the antisymmetry
of the wave function would require radial excitations and/or higher
orbital angular momentum in the $S_h=0$ component. However,
in the $bc$ sector, the mass eigenstate $\Xi$ $(\Omega)$ particles are
 mixtures of the $\Xi_{bc},\,\Xi'_{bc}$
($\Omega_{bc},\,\Omega'_{bc}$) states listed in
Table~\ref{tab:hh-baryons}. Indeed, the mixing angle is large,  around
30 $\deg$ (\citep{Roberts:2007ni, Albertus:2009ww}). This hyperfine
mixing greatly affects the decay widths of doubly heavy baryons
involving $\Xi_{bc}-$baryons. This was firstly established by Roberts and
Pervin~\cite{Roberts:2009} and later on confirmed in
Ref.~\citep{Albertus:2009ww}. Nevertheless, the HQSS predictions for
the weak matrix elements of the unmixed states derived in
Ref.~\citep{Flynn:2007qt} can be used to predict those of the mixed
states, and moreover they might be used in the future to
 experimentally extract information on the mixtures in the actual
 physical $bc-$baryon states~\cite{Albertus:2009ww}.

\section{Spin Symmetry}
The invariance under separate spin rotations of the $b$ and $c$ quarks
leads to relations between the form factors for vector and
axial-vector current decays of  cascade $bc$ baryons to cascade
$cc$ baryons. These decays are induced by the weak $b \to c \,l^-
\nu_l$ ($l=e,\mu$) transition. To represent the lowest-lying $L=0$ 
$bcq$ baryons we will use wavefunctions comprising tensor products of
Dirac matrices and spinors, namely:
\begin{eqnarray}
\label{eq:Bprimebc}
B'_{bc} &= &
 -\left[\frac{(1+\vsl)}2 \gamma_5\right]_{\alpha\beta} u_\gamma(v,r)\\
\label{eq:Bbc}
B_{bc} &= &
 \left[\frac{(1+\vsl)}2 \gamma_\mu\right]_{\alpha\beta} \left[\frac1{\sqrt3}
 (v^\mu+\gamma^\mu)\gamma_5  u(v,r)\right]_\gamma\\
\label{eq:Bstarbc}
B^*_{bc} &=& \Xi^*_{bc} = 
 \left[\frac{(1+\vsl)}2 \gamma_\mu\right]_{\alpha\beta} u^\mu_\gamma(v,r)
\end{eqnarray} 
where we have indicated Dirac indices $\alpha$, $\beta$ and $\gamma$
explicitly on the right-hand sides and $r$ is a helicity label for the
baryon\footnote{We use the standard relativistic normalization for
  hadronic states and our spinors satisfy $\bar u u = 2 m$, $\bar
  u^\mu u_\mu = - 2 m$ where $m$ is the mass of the state.}. For the
$B^*_{bc}$, $u^\mu_\gamma(v,r)$ is a Rarita-Schwinger spinor. These
wavefunctions can be considered as matrix elements of the form
$\langle0 | c_\alpha \bar{q^c}_\beta b_\gamma |
B^{(\prime*)}_{bc}\rangle$ where $\bar{q^c}=q^T C$ with $C$ the
charge-conjugation matrix. We couple the $c$ quark and light quark to
spin $0$ for the $B'_{bc}$ or $1$ for the $B_{bc}$ and $B^*_{bc}$
states. Under a Lorentz transformation, $\Lambda$, and $b$ and $c$
quark spin transformations $S_b$ and $S_c$, a wavefunction of the form
$\Gamma_{\alpha\beta}\, u_\gamma$ transforms as:
\begin{equation}
\label{eq:spintransfs}
\Gamma\,u \to S(\Lambda) \Gamma S^{-1}(\Lambda)\; S(\Lambda)u,
\quad
\Gamma\,u \to S_c \Gamma \, S_b u.
\end{equation}
The states in Eqs.~\eqref{eq:Bprimebc}, \eqref{eq:Bbc} and
\eqref{eq:Bstarbc} have a common normalization $\bar u u
\mathrm{Tr}(\Gamma \overline\Gamma)$ and are mutually orthogonal. To
build states where the $b$ and $c$ quarks are coupled to definite
spin, we need the linear combinations
\begin{eqnarray}
\label{eq:Sbceq0}
|0;1/2,M\rangle_{bc} &=& -\frac12 |0;1/2,M\rangle_{cq}
 + \frac{\sqrt3}2 |1;1/2,M\rangle_{cq}\nonumber\\
\label{eq:Sbceq1}
|1;1/2,M\rangle_{bc} &=& \frac{\sqrt3}2 |0;1/2,M\rangle_{cq}
 + \frac12 |1;1/2,M\rangle_{cq}\nonumber\\
|1;3/2,M\rangle_{bc} &=& |1;3/2,M\rangle_{cq}
\end{eqnarray}
where the second and third arguments are the total spin quantum
numbers of the baryon and the first argument denotes the total spin of
the $bc$ or $cq$ subsystem.
For the $cc$ baryons there are some differences because we have
two identical quarks. In this case the states are:
\begin{eqnarray}
\label{eq:Bprimecc}
B'_{cc} &=&
 -\sqrt{\frac23}
 \left[\frac{(1+\vsl)}2 \gamma_5\right]_{\alpha\beta} u_\gamma(v,r)\\
\label{eq:Bcc}
B_{cc} &=&
 \left[\frac{(1+\vsl)}{\sqrt{2}} \gamma_\mu\right]_{\alpha\beta}
 \left[\frac1{\sqrt3}(v^\mu+\gamma^\mu)\gamma_5  u(v,r)\right]_\gamma\\
\label{eq:Bstarcc}
B^*_{cc} &=& \Xi^*_{cc} =
 \sqrt{\frac12}
 \left[\frac{(1+\vsl)}2 \gamma_\mu\right]_{\alpha\beta} u^\mu_\gamma(v,r)
\end{eqnarray}
The two charm quarks can only be in a symmetric spin-$1$ state and
therefore $B'_{cc}$ and $B_{cc}$ correspond to the same baryon state
$\Xi_{cc}$ (or $\Omega_{cc}$ if the light quark is $s$). We can now
construct amplitudes for semileptonic cascade $bc$ to cascade $cc$
baryon decays, determined by matrix elements of the weak current
$J^\mu = \bar c \gamma^\mu(1-\gamma_5) b$. We first build transition
amplitudes between the $B^{(\prime*)}_{bc}$ and $\Xi^{(*)}_{cc}$
states and subsequently take linear combinations to obtain transitions
from $\Xi^{(\prime*)}_{bc}$ states. The most general form for the
matrix element respecting the heavy quark spin symmetry is
\begin{multline}
\label{eq:ME}
\langle\Xi^{(*)}_{cc},v,k,M'|J^\mu(0)|B^{(\prime*)}_{bc},v,M\rangle \\
\begin{aligned}
 &=\bar u_{cc}(v,k,M') \gamma^\mu(1-\gamma_5) u_{bc}(v,M)
   \mathrm{Tr}[\Gamma_{bc}\Omega \overline\Gamma_{cc}]\\ 
 & \quad + \bar u_{cc}(v,k,M')
      \Gamma_{bc}\Omega\overline\Gamma_{cc} \gamma^\mu(1-\gamma_5) u_{bc}(v,M)
\end{aligned}
\end{multline}
where $M$ and $M'$ are the helicities of the initial and final
states  and
$\Omega = -\eta(\omega)/2$, with $\omega=v\cdot v'$.  To simplify, we 
use  the equations of motion
($\vsl u=u$, $\vsl\Gamma=\Gamma$, $\gamma_\mu u^\mu=0$, $v_\mu
u^\mu=0$), while terms with $\slashchar k$ will always lead to
contributions proportional to $v\cdot k$ which is set to $0$ at the
order we are working. We also make use of the relations $\bar u
\gamma_\mu u = \bar u v_\mu u$, $\bar u \gamma_5 u = 0$, $\bar u
\slashchar k u =0$ and $\bar u \slashchar k \gamma_\mu \gamma_5 u = -
\bar u \slashchar k v_\mu \gamma_5 u$. Our results for cascade $bc$ to
cascade $cc$ transition matrix elements are\cite{Flynn:2007qt}:
\begin{align}
\label{eq:XibctoXicc}
\Xi_{bc}\to \Xi_{cc} \qquad &
 \eta\, \frac{1}{\sqrt2}\bar u_{cc}\left(2\gamma^\mu
 -\frac43\gamma^\mu\gamma_5\right) u_{bc}\\
\Xi'_{bc}\to \Xi_{cc} \qquad &
 -\sqrt{\frac23} \eta\, \bar u_{cc}\left(-\gamma^\mu\gamma_5\right) u_{bc}\\
\Xi_{bc}\to \Xi^*_{cc} \qquad &
 -\sqrt{\frac23} \eta\, \bar u^\mu_{cc} u_{bc}\\
\Xi'_{bc}\to \Xi^*_{cc} \qquad &
 -\sqrt2 \eta\, \bar u^\mu_{cc} u_{bc}\\
\label{eq:XibcstartoXicc}
\Xi^*_{bc}\to \Xi_{cc} \qquad &
 -\sqrt{\frac23} \eta\, \bar u_{cc} u^\mu_{bc}\\
\Xi^*_{bc}\to \Xi^*_{cc} \qquad &
 -\sqrt2 \eta\, \bar u^\lambda_{cc}\left(\gamma^\mu-\gamma^\mu\gamma_5\right)
 u_{bc\,\lambda}\label{eq:eqb}
\end{align}
If the $b$ and $c$ quarks become degenerate, then vector current
conservation ensures that $\eta(1)=1$. Similarly, relations for the
decays of $bb$ baryons to $bc$ baryons can be obtained~\cite{Albertus:2009ww}. 
\end{multicols}
\ruleup
\begin{center}
%\vspace{0.4cm}
\makebox[0pt]{\includegraphics[scale=0.275]{Xibccc_all.eps}\hspace{0.65cm}\includegraphics[scale=0.275]{Xibbbc_all.eps}} 
%\vspace{0.24cm}
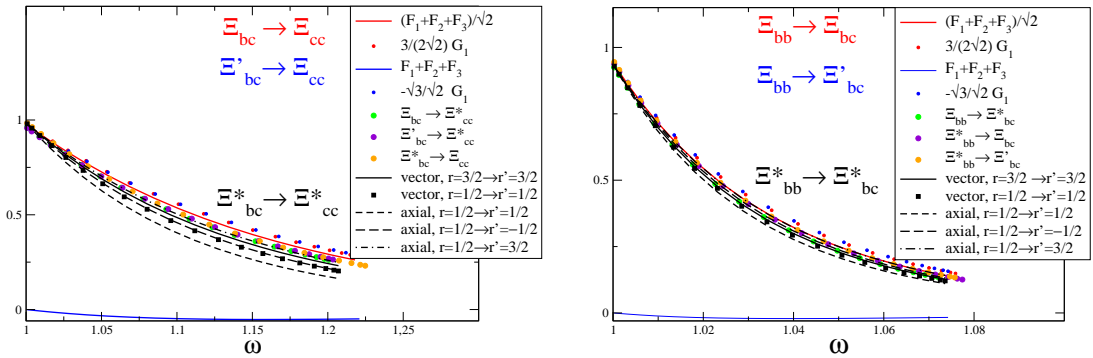
\figcaption{Left panel: Different $\eta$ functions obtained for
$\Xi_{bc}^*\to\Xi_{cc}^*$ transitions (black curves) using the vector
or the axial part of the weak transition current, and for different
spin configurations. We also show the
corresponding results obtained for $1/2\to1/2$ and
$1/2\longleftrightarrow3/2$ transitions. Form factors are taken from
Refs.~\citep{Albertus:2006ya,Hernandez:2007qv}. Baryon
  wave functions are obtained by means of a variational
  approach~\protect\cite{Albertus:2001pb,Albertus:2003sx,Albertus:2006ya},
  while the semileptonic decay widths are computed in coordinate
  space~\protect\cite{Albertus:2006ya} by using a scheme derived in
  Ref.~\citep{Albertus:2004wj}. Right
panel: same as left panel for $bb\to bc$ transitions. \label{fig:fig1}} 
\end{center}
\ruledown
\begin{multicols}{2}
\section{Results and conclusions}
All hadronic matrix elements of the $J=V-A$ current implicit in the
left hand sides of Eqs.~(\ref{eq:XibctoXicc})--(\ref{eq:eqb}), near
zero recoil, are given in terms of a unique function, $\eta(\omega)$,
of the product of four velocities, up to corrections suppressed by the
mass of the charm and bottom quarks. These matrix elements are usually
parameterized in terms of form factors, whose number is restricted by
Lorentz covariance and the discrete $C,P,T$ symmetries. There are six
form factors to describe $\Xi_{bc}\to\Xi_{cc}$, another six for
$\Xi'_{bc}\to\Xi_{cc}$, eight each for $\Xi_{bc}\to\Xi^*_{cc}$,
$\Xi'_{bc}\to\Xi^*_{cc}$ and $\Xi^*_{bc}\to\Xi_{cc}$, and even
more~\cite{Faessler:2009xn} for $\Xi^*_{bc}\to\Xi^*_{cc}$. In
Fig.~\ref{fig:fig1}, we show constituent quark model results for the
various form factors~\cite{Albertus:2006ya,Hernandez:2007qv}.  We see,
that to a good approximation, better in the $bb\to bc$ case as one is
closer to the infinite heavy quark mass limit, all $1/2\to1/2$,
$1/2\longleftrightarrow 3/2$ and $3/2\to3/2$ transitions are governed
in terms of just one function, as deduced in
Eqs.~(\ref{eq:XibctoXicc})--(\ref{eq:eqb}) for the $bc\to cc$
transitions.  This function is different for the $bc\to cc$ and $bb\to
bc$ cases due to heavy flavour symmetry breaking.

To the extent that one is close enough to the infinite heavy quark
mass limit and near zero recoil, we can make use of the HQSS results
in Eqs.(\ref{eq:XibctoXicc})--(\ref{eq:eqb}) and the similar ones for
$bb\to bc$ transitions, to approximate the hadron tensor that governs
these decays. Thus, it is possible to construct ratios of widths 
where the dependence on the universal $\eta(\omega)$ function will
cancel out, in the strict near zero recoil approximation (for details,
see Ref.~\citep{Hernandez:2007qv}). In Table~\ref{tab:hqsr} we show
different model predictions for several ratios that should be one in
the infinitely heavy quark limit. We see that the calculations by
Hern\'andez {\it et al.}~\cite{Hernandez:2007qv}, Ebert {\it et
  al.}~\cite{Ebert:2004ck} and Faessler {\it et
  al.}~\cite{Faessler:2009xn} turn out to be in reasonable agreement
with HQSS predictions. Only the second of the ratios can be computed
from the results of Roberts and Pervin in \citep{Roberts:2009}, and we
find  a value of 0.80 (0.88) for $\Xi$ ($\Omega$) type
baryons. The  results in
Ref.~\citep{Sanchis-Lozano:1994vh} are also not inconsistent 
with HQSS constraints. However, HQSS predictions turn out to be
incompatible with the results of Ref.~\citep{Guo:1998yj}, hinting
at   problems  in the model or  the
calculation in that work.

\end{multicols}
\begin{center}
\tabcaption{Decay width ratios for semileptonic $bb\to bc$ decay of
  doubly heavy $\Xi$ and $\Omega$ baryons. 
In all cases the  approximate result 
  obtained using HQSS is 1.\label{tab:hqsr}}
\begin{tabular*}{170mm}{@{\extracolsep{\fill}}c|cccccccc}
\toprule
&\multicolumn{2}{c}{\citep{Hernandez:2007qv}
}&\multicolumn{2}{c}{\citep{Ebert:2004ck} }
&\multicolumn{2}{c}{\citep{Guo:1998yj}}&
\multicolumn{2}{c}{\citep{Faessler:2009xn}}\\
\multicolumn{1}{c|}{\Large $bb\to
    bc$} &$\Xi$ &$\Omega$ &$\Xi$ &$\Omega$ &$\Xi$ &$\Omega$ &$\Xi$ &$\Omega$ \\\hline
&&&&&\\
$\frac{\Gamma(B^*_{bb}\to  B'_{bc}\,l\bar\nu_l)}{3\,\Gamma( B^*_{bb}\to  B_{bc}\,l\bar\nu_l)}$&$1.00^{+0.01}_{-0.04}$&
$1.00^{+0.03}_{-0.01}$&0.99&0.99&0.05&--- &$0.9^{+0.5}_{-0.3}$ & $0.9^{+0.6}_{-0.4}$\\
 &&&&&\\
 $\frac{\Gamma({ B}_{bb}\to { B}^*_{bc}\,l\bar\nu_l)}{\frac23\,\Gamma({ B}_{bb}\to
 B'_{bc}\,l\bar\nu_l)
}$&$0.86^{+0.08}_{-0.06}$&$0.86^{+0.05}$&0.96&0.99&
  9.53&---&$0.9^{+0.5}_{-0.3}$ & $0.9^{+0.5}_{-0.3}$ \\
 &&&&&\\
 $\frac{\Gamma({ B}^*_{bb}\to { B}_{bc}\,l\bar\nu_l)}{\frac13\,\Gamma({ B}_{bb}\to{ B}'_{bc}\,l\bar\nu_l)}$&$0.98^{+0.09}_{-0.03}$&
$0.97^{+0.06}_{-0.14}$&1.01&1.03&36.4&---
&$1.0^{+0.5}_{-0.3}$ & $0.9^{+0.5}_{-0.4}$ \\
 &&&&&\\
 $\frac{\Gamma({ B}^*_{bb}\to { B}^*_{bc}\,l\bar\nu_l)}{\Gamma({ B}_{bb}\to
{ B}_{bc}\,l\bar\nu_l)+\frac12\,\Gamma({ B}_{bb}\to {
  B}^*_{bc}\,l\bar\nu_l)}$&$0.94^{+0.07}_{-0.06}$&$0.93^{+0.11}_{-0.10}$&1.01&1.01&
  0.31&---&$1.1^{+0.8}_{-0.5}$ & $1.1^{+0.8}_{-0.5}$\\
\bottomrule
\end{tabular*}
\end{center}
\begin{multicols}{2} 

\end{multicols}

\vspace{-2mm}
\centerline{\rule{80mm}{0.1pt}}
\vspace{2mm}

\begin{multicols}{2}

\end{multicols}

\clearpage

\end{document}